\documentclass[journal=jacsat,manuscript=article]{achemso}
\usepackage{chemformula} 
\usepackage[T1]{fontenc} 
\usepackage{graphicx}
\usepackage{amsmath}
\newcommand{\angstrom}{\text{\normalfont\AA}}
\usepackage{amssymb}
\usepackage{bm}
\usepackage{caption}
\usepackage{subcaption}
\usepackage{hyperref}
\usepackage{nameref}
\usepackage{booktabs}
\usepackage{multirow}
\usepackage{wrapfig}
\usepackage{comment}

\hypersetup{pdfborder=0 0 0,colorlinks=true,citecolor=black,linkcolor=black}
\input epsf

\newcommand\THEOSMARVEL{Theory and Simulation of Materials (THEOS) and National Centre for Computational Design and Discovery of Novel Materials (MARVEL), {\'E}cole Polytechnique F{\'e}d{\'e}rale de Lausanne, 1015 Lausanne, Switzerland}
\newcommand\UNIBI{Universit{\`a} degli studi di Milano Bicocca, Piazza dell'Ateneo Nuovo 1, 20126 Milano, Italy}
\newcommand\BIQUTE{Bicocca Quantum Technologies (BiQuTe) Centre, I-20126 Milano, Italy}
\newcommand\PSI{Laboratory for Materials Simulations (LMS), Paul Scherrer Institut, CH-5232 Villigen PSI, Switzerland}
\newcommand\BREMEN{U Bremen Excellence Chair, Bremen Center for Computational Materials 
Science, and MAPEX Center for Materials and Processes, Universit{\"a}t Bremen, 28359 Bremen, Germany}

\author{Chiara Cignarella}
\affiliation{\BREMEN}
\alsoaffiliation{\THEOSMARVEL}
\author{Nicola Marzari}
\affiliation{\THEOSMARVEL}
\alsoaffiliation{\PSI}
\alsoaffiliation{\BREMEN}
\author{Davide Campi}
\email{davide.campi@unimib.it}
\affiliation{\THEOSMARVEL}
\alsoaffiliation{\UNIBI}
\alsoaffiliation{\BIQUTE}

\title{High-throughput discovery for one-dimensional exfoliable materials}

\begin{document}
\begin{abstract}
The isolation of single or few-layer thick two-dimensional materials has led to remarkable fundamental discoveries and revolutionary applications, greatly pushing our knowledge of condensed matter physics as well as the development of ultra-scaled highly-efficient devices. A further reduction in dimensionality, \emph{i.e.} the isolation of single or few-atom  wires would give us easier access to fascinating quantum phenomena and novel extraordinary applications. 
Experimental exfoliation of one-dimensional wires from weakly bonded three-dimensional crystals has attracted increasing interest in recent years, owing to the wire's extreme electronic properties and potential for device applications. 
To boost the research and extend the range of potentially exfoliable wires, here we report an high-throughput screening aimed at identifying one-dimensional materials that can be extracted from weakly bonded, and experimentally reported, three-dimensional parents. 
In this work, we present a database of 686 candidate wires, for which we have introduced and evaluated three physically motivated exfoliation criteria to better reflect the likelihood and quality of exfoliation from the three-dimensional compounds.
Moreover, we provide a comprehensive characterization of their electronic and vibrational properties.
In particular, we evaluate the full phonon spectrum along the periodic direction for each of the candidates to identify possible dynamical instabilities, to which one-dimensional materials are particularly susceptible.
The database is openly available for a full exploration of the 1D materials, their electronical and dynamical properties, and the provenance of their 3D parent, and it will provide a valuable platform for exploring materials at the ultimate limit of dimensionality.
\end{abstract}
\section{Introduction}
In the past twenty years the progresses in the isolation of two-dimensional materials brought great fundamental advances in condensed matter physics as well as applications with enormous potential in different fields from electronics and optoelectronics\cite{Radisavljevic:2011,Wang2012,Liu:2014, Roy2014, Chhowalla2016, Klinkert2020} to  electro- and photo-catalysis\cite{He2012,Sun2012} to the exploitation of exotic states of matter\cite{Qian2016,Marrazzo2018,Campi2021}. The possibility to descend a further step in dimensionality, isolating or growing reproducible single or few-atom one-dimensional (1D) wires can unlock an easier exploration of rare physical phenomena like charge-density waves\cite{gruner2018density,peierls1955quantum,pouget2016peierls,giamarchi2016one}, Luttinger liquids\cite{luttinger1963exactly,haldane1981luttinger,voit1995one,bechstedt2026antiscreening}, or exciton condensates\cite{Varsano2017}  
and allow the realization of electronic devices reaching the ultimate scaling and performance limits \cite{BALANDIN202274,Geremew2018,Ott2019,Park2020,zhu2021spectrum,cignarella2024searching}.
However, while the field of two-dimensional materials is, after years of development, in quite mature stage, with tens of monolayers that can be routinely isolated, the field of one-dimensional materials is comparatively still in its infancy.  Over the years various techniques have been developed to grow one-dimensional chains on surfaces from bottom-up approaches \citep{gambardella2002ferromagnetism,crain2004chains,zeng2008charge,ferstl2016self,qin2018epitaxial,guo2022direct} 
but in many cases the resulting system strongly interacts with the environment diluting its 1D character\citep{sanna2018one,yogi2022electronic}. As for freestanding materials, carbon nanotubes have been the object of decades of research but the difficulties in consistently controlling their properties made their use    
difficult particularly for nanoelectronics applications\citep{iijima1991helical,iijima1993single,ajayan1997nanometre,tans1997individual,kim2001thermal}. 
The need for new candidates for easily reproducible,  one-dimensional freestanding materials sparked interest towards crystals in which strongly bonded atomic wires are held together by weak Van-der-Waals interactions\citep{BALANDIN202274,meng2022one,balandin2022one} in which single wires could potentially be isolated from the bulk. 
Since the first successful mechanical and liquid exfoliations of 
one-dimensional transition metal trichalcogenides\citep{stolyarov2016breakdown,island2017electronics,lipatov2018quasi,geremew2018current} the field has grown rapidly, with several quasi-1D van der Waals materials synthesized in the last few years. Examples include Nb$_2$Se$_9$ \cite{oh2018inorganic,kim2018mechanical}, the Ta$_2$X$_3$Se$_8$ family \cite{liu2016direct,jeong2022ternary,jeong2021one,antipina2025unlocking,choi2024scalable}, HfSnS$_3$\cite{lu2023synthesis}, HfS$_3$\cite{lipatov2023electronic}, Ta$_2$Se$_8$I$_2$\cite{teeter2024achieving,ghosh2023low}, among others \cite{choi2022one,yang2021observation,mahdikhany2024ultra,qu2021ultrathin,zhao2026extreme}.
The experimental exfoliation of MoI$_3$ \cite{teeter2024achieving} has further paved the way to achieving the single-chain limit.

The phase space of quasi-1D systems likely expands far beyond the set of materials experimentally studied so far. 
In this regard, high-throughput (HT) computational studies can significantly expand the landscape of atomic wires, as already demonstrated for two-dimensional materials \cite{mounet2018two,haastrup2018computational,Zhou2019,campi2023,zolotarev2026}. 
Extensions of data-mining algorithms developed for 2D materials \cite{cheon2017data,larsen2019definition} represented early efforts in this direction. A dedicated high-throughput study by Moustafa et al \cite{moustafa2022computational} identified 288 one-dimensional systems weakly bonded to existing three-dimensional bulks using the scoring parameter developed in Ref. \citenum{larsen2019definition}, and further expanded this subset through chemical substitution. 
Similar HT searches of the ICSD database, using either data-mining \cite{lu2021high,cheon2017data} or graph-theory-based classification \cite{Li2022}, identified 367 and 244 systems, respectively.
Other high-throughput studies have focused on
specific families of one-dimensional van der Waals materials, such as transition metal chalcogenides \cite{shang2020atomic}, metal halide nanowires \cite{fu2022transition}, for specific properties such as structural or kinetical stability \cite{zhu2021spectrum,zong2026van}.
Larger datasets of theoretically predicted atomic wire materials have been created, but these rely on elemental substitution or generative machine-learning algorithms to extend the composition space of 1D wires \cite{Moustafa2023,zhu2023machine,zhao2026extreme}.
In this work, we aim to uncover plausible candidates that could be exfoliated into ultrathin nanowires starting from existing parent compounds. To do that, we perform an extensive high-throughput screening of experimentally known compounds taken from three different databases to identify 1D systems that could be isolated from three-dimensional crystals. Importantly, we start from databases of three-dimensional crystals, the ICSD ~\cite{ICSD}, COD~\cite{COD} and MPDS \cite{Villars1998,mpds}, and select only materials with a reported experimental synthesis.
Since the little experimental data on the exfoliation of 1D systems available to date are insufficient to properly define a precise exfoliation criterion, we here propose three descriptors for the exfoliation, ranking all the materials accordingly. We define three physically-grounded criteria, that could be directly comparable to experiments, accounting for the thickness of the atomic wires as well as their stiffness to describe their ease of exfoliation. 
We optimize the geometry of the extracted 1D-systems as isolated in vacuum, and systematically characterize their electronic properties, including band structures with and without spin–orbit coupling and charge-carrier effective masses.
Crucially, we evaluate for the first time their dynamical stability by computing the full phonon dispersion over a uniform grid. 
The resulting \href{}{Materials Cloud One-Dimensional Structure Database (MC1D)} is openly available and contains 686 unique one-dimensional wires, providing electronic properties and phonon band dispersions for each wire. To our knowledge, the MC1D constitutes the largest database to date of 1D wires derived exclusively from experimentally reported three-dimensional compounds.

\section{Results and discussions}
\subsection{Identification of atomic wires in bulk materials}
\begin{figure}[h!]
    \centering
    \includegraphics[scale=0.5]{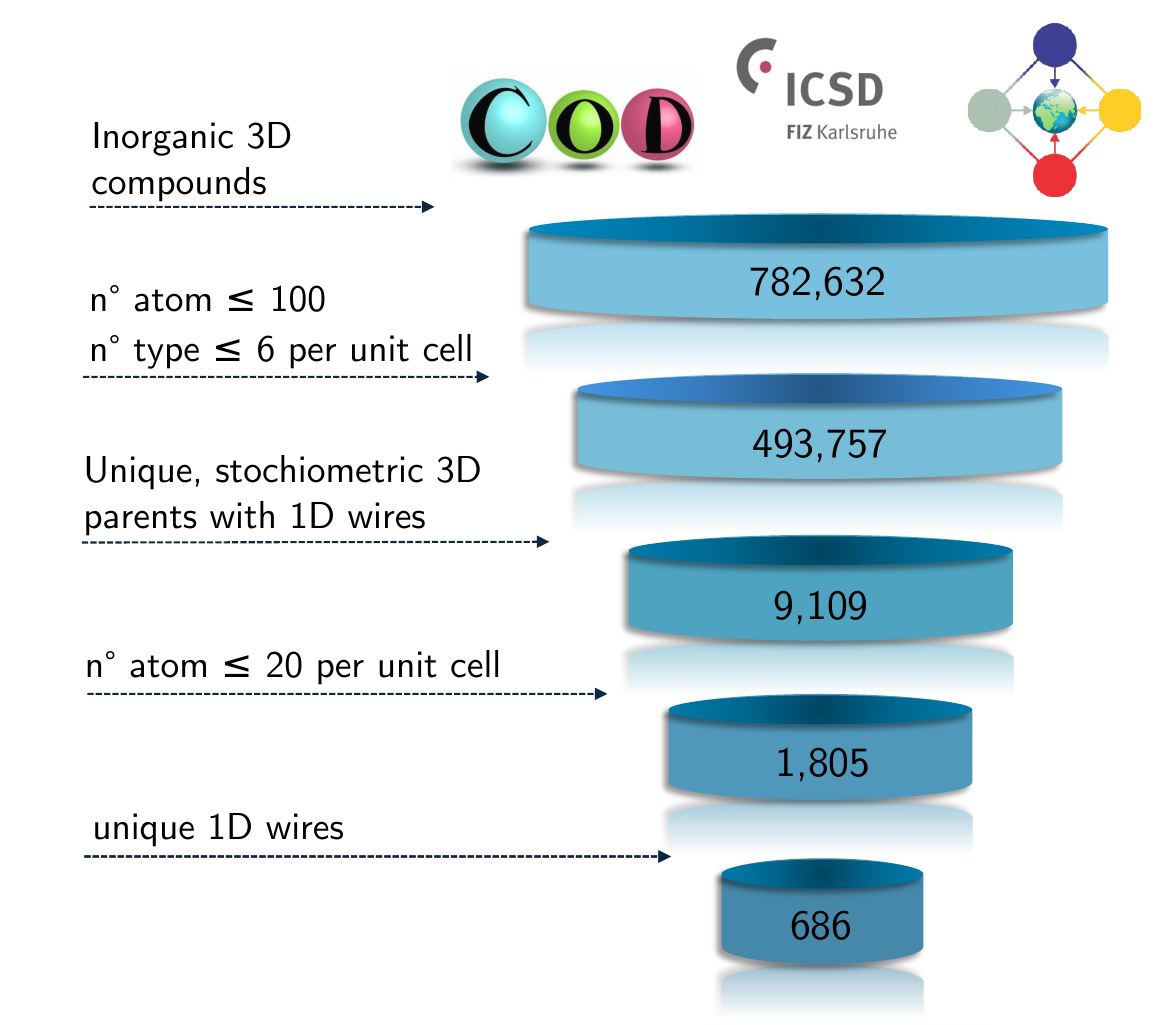}
    \caption{Schematic portrayal of the high-throughput screening for one-dimensional exfoliable materials.}
    \label{funnel}
\end{figure}

In this work we follow a recipe similar to the one detailed in Refs.\citenum{Mounet2018,campi2023} for two-dimensional materials, properly adapted to cope with 1D-systems. We start by extracting the bulk 3D crystal structures, in the form of CIF files~\cite{CIF}, from three experimental repositories: ICSD~\cite{ICSD} ($147731$ structural entries), COD~\cite{COD} ($279885$ entries) and the Pauling File (MPDS) \cite{Villars1998,mpds} ($355016$ entries). 
After excluding structures with partial occupations, structures with implicit hydrogen, structures explicitly reported as purely theoretical in the cif file, or obviously wrong structures, we imported them into AiiDA using pymatgen~\cite{pymatgen}.
The screening workflow is schematically represented in Figure \ref{funnel}.

We then analyze all the structures with up to 100 atoms per unit cell with a geometrical screening procedure already used for two-dimensional materials in Refs.\citenum{Mounet2018,campi2023}, which consists in building a supercell, accounting for the connection between two atoms when their distance is smaller than the sum of their respective VdW radii at least by $\Delta$ (taken between $1.0~\angstrom $ and $1.5~\angstrom$)\cite{alvarez} and computing the vectors connecting an atom with its periodic replicas. We then select materials for which the dimension of the subspace created by any of these vector ensembles is one, indicating the presence of one-dimensional substructure.

This geometrical selection identified a total of 9.109 materials with vdW-bonded 1D substructures.
These selected structures 
are then processed with the spglib software~\cite{spglib} to find the primitive cells, and filtered for uniqueness using the pymatgen structure matcher\cite{CMPZ}. 

For all these structures we optimized the cell parameters and internal geometry at their theoretical equilibrium values employing the non-local vdW-compliant functionals vdW-DF2 functional~\cite{lee2010} with C09 exchange~\cite{cooper2010,hamada2010} (DF2-C09). These optimized structures are the basis for computing the exfoliation descriptors described in the following section.  

Crystallographic databases (ICSD, COD and Pauling File) usually flag crystal structures in which hydrogen positions could not be reliably determined by X-ray diffraction. However, these flags are not always reliable, and some incomplete structures can remain unflagged and thus slip through our automatic filters. We therefore manually inspected all the resulting 1D wires and kept in the final database only those derived from complete crystal structures. After this final step, the database contains 686 unique wires.

\begin{figure}[h!]
    \centering
    \begin{subfigure}[c]{0.5\textwidth}
     \includegraphics[scale=0.75]{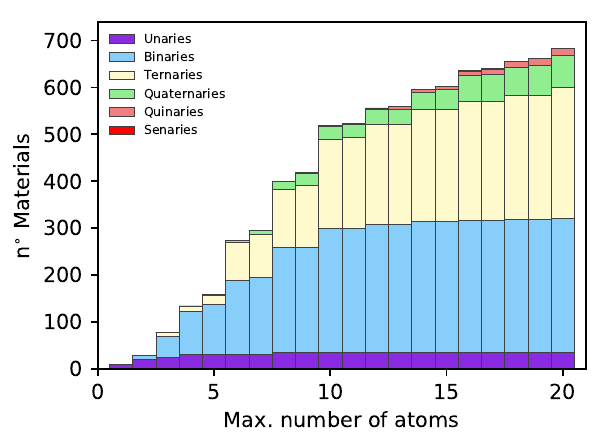}
    \caption{}\label{nadist}
    \end{subfigure}%
        \begin{subfigure}[c]{0.5\textwidth}
  \includegraphics[scale=0.24]{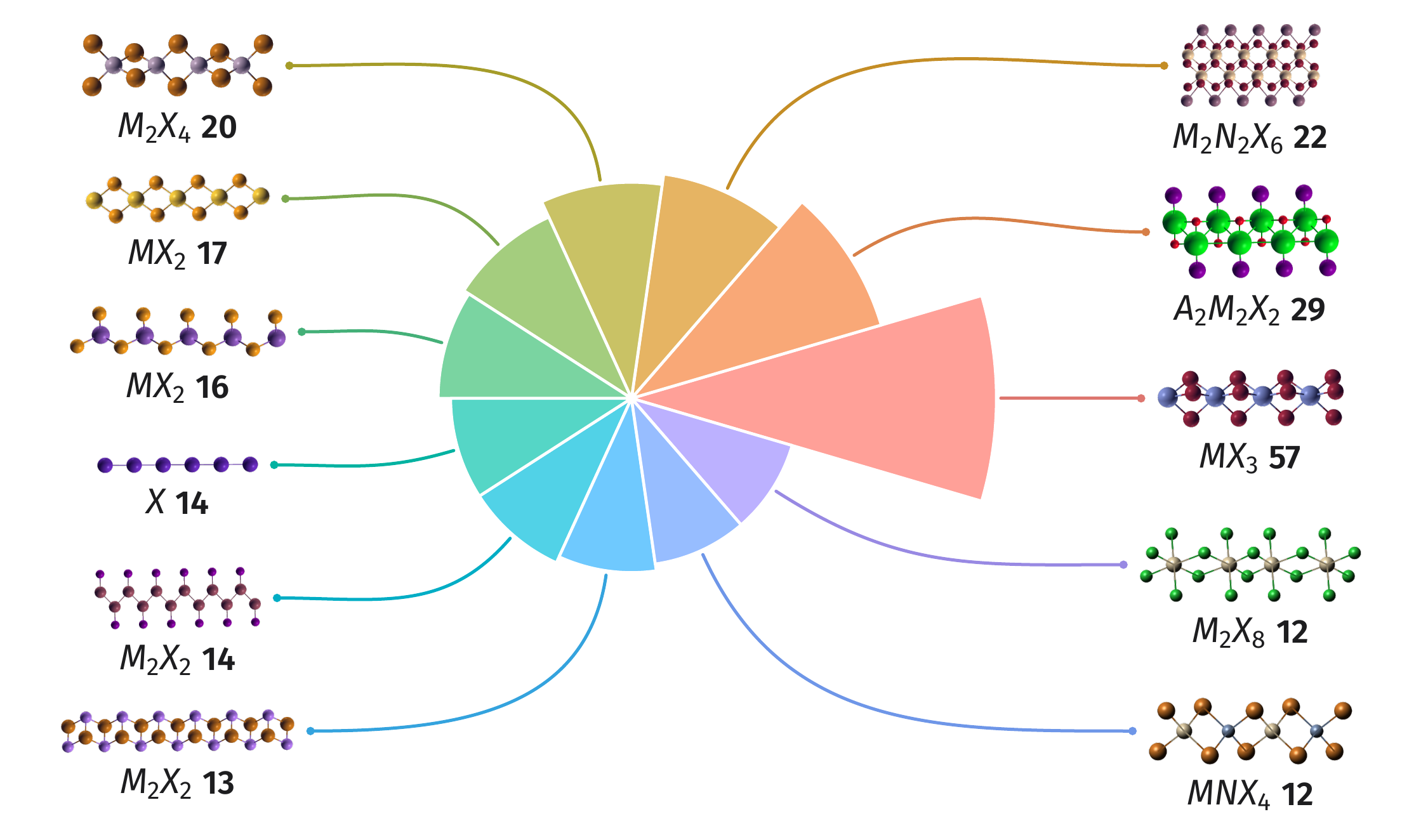} 
    \caption{}\label{1Dprototypes}
    \end{subfigure}
    \caption{(a) Number of structures as a function of the number of atoms in the bulk primitive cell. Binary and ternary compounds dominate the database of selected structures while mono-elemental compounds are rare as they are materials with more complex stoichiometries. (b) Distribution of the ten most common 1D prototypes accounting for a total of 226 materials and a schematic representation of each structural prototype. The most common prototype by far (accounting for 57 structures) is the MX$_3$, where usually M is a transition metal and X a chalcogen or an halogen. }
\end{figure}

In Fig. \ref{nadist} we report the distribution of atomic types as a function of the number of atoms in the bulk parent materials. We can notice that the database is largely dominated by binary and ternary compounds; the number of binary compounds is proportionally larger than for two-dimensional materials\cite{Mounet2018} while the contribution of materials with more than four atoms is lower, hinting towards generally simpler stoichiometries in one-dimensional materials. Mono-elemental materials are generally rare and dominated by exotic forms of group-V and group-6 elements.  Most of the wires come from bulk with relatively low symmetry (P2$_{1}$, P-1 and P2/m symmetries) as reported in Fig. S1 of the Supporting Information.
To provide a more direct overview of our database, we classify the 1D materials into different prototypes, according to their structural similarity, distinguishing the positions occupied by different atomic types but ignoring the precise chemical composition.  In out database we find a total of $320$ prototypes picturing a less complex landscape than for two-dimensional materials, where the structural prototypes were more than one thousand\cite{campi2023}. The eleven most common ones are reported in Fig.~\ref{1Dprototypes}. The dominant prototype by far (with 57 materials) is MX$_3$ compounds, with M typically being a transition metal and X a chalcogen or an halogen atom. Surprisingly the second most common prototype is a more complex ternary oxide with a total 29 structures. The third  and fourth places are taken by ternary M$_2$N$_2$X$_6$ compounds and binary combination in two-to-one ratio of a chalcogen or halogen with a transition metal. The ideal monoatomic chain takes only the seventh position with just 14 compounds showing this arrangement. 

\subsection{Exfoliation Criteria}
The definition of a single and reliable parameter for the exfoliability likelihood for one-dimensional materials is not straightforward. In our previous works on two-dimensional materials\cite{Mounet2018,campi2023} we based our analysis on the binding energy per area unit. Thanks to the abundance of experimental data on the exfoliation of two-dimensional materials, this descriptor can be thoroughly benchmarked and a cutoff on its value can be sensibly taken considering the upper limit of the computed binding energies of materials that have been experimentally isolated down to the monolayer limit. For one-dimensional materials however, experimental data are scarce and definitely insufficient to clearly determine an exfoliation criteria or to estimate for it a reasonable threshold for exfoliability.  For this reason, we propose here three different descriptors and we report all the extracted one-dimensional materials without imposing an hard cutoff on any of these descriptors. The first descriptor is the simplest extension of the criteria used for two-dimensional materials, \emph{i.e.} the binding energy per unit of length, computed as the energy difference between the optimized three-dimensional bulk and the sum of the energies of the isolated subsystems, divided by the length of the wire unit cell along the periodic direction. This approach has been used in previous works\citep{shang2020atomic,zhu2021spectrum, moustafa2022computational,Li2022} however it is rigorously applicable only to idealized 1D chains with a negligible thickness. This might lead to an improper estimation of the exfoliation likelihood for thicker structures that deviates from the ideal dimensionless atomic chain. To overcome this problem and define a quantity which is closer to the one used for two-dimensional materials we also defined an exfoliation energy per unit of area. We define the area as the lateral surface of a cylinder with the height corresponding to the periodic direction of the cell and the volume computed using the quantum volume defined in Ref. \citenum{cococcioni2005electronic}.
Finally it is reasonable to assume that also the strength of the quasi-1D structure along its periodic direction would play a role in determining the quality of the isolated wire, as the in-plane Young modulus plays a role in determining the quality of the flakes of two-dimensional materials obtained from layered structures \cite{Backes2019}. To account for this we defined a third metric as the ratio between the binding energy per area unit as defined above and the one dimensional Young modulus. The lower the ratio, the easier it should be to obtain high-quality long wires during the exfoliation process.   

\begin{figure}[h!]
\includegraphics[scale=1.0]{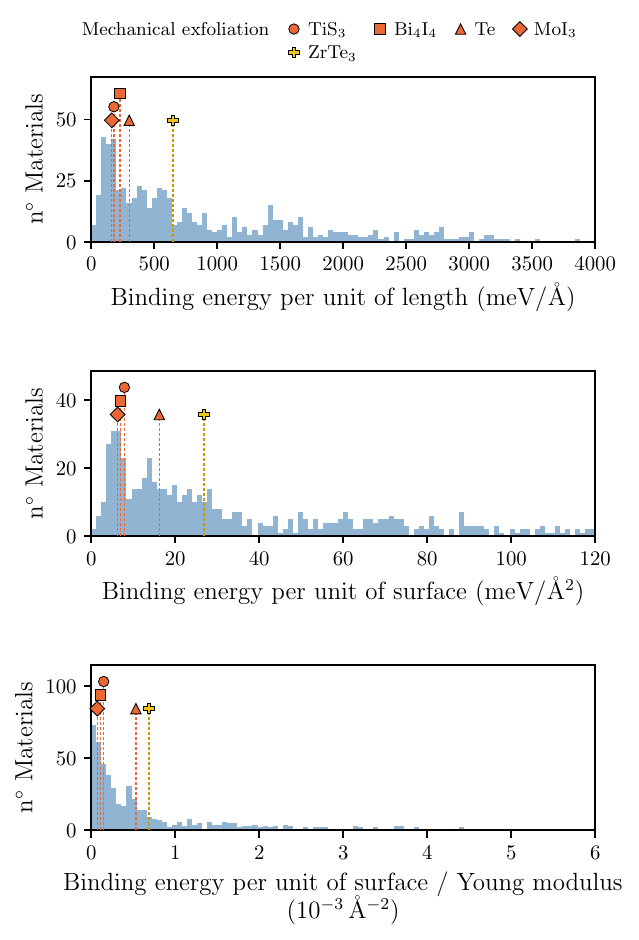}
\caption {(a) Distribution of the binding energies per unit of length. (b) Distribution of the binding energies per unit of surface, exploiting the quantum volume definition. (c) Distribution of the binding energies per unit of surface exploiting the quantum volume definition divided by the 1D Young modulus. We highlight the binding energies according to each measure of known mechanically exfoliated wires:  TiS$_3$\cite{island2015tise3,lipatov2018quasi}, MoI$_3$ \cite{teeter2024achieving}, Bi$_4$I$_4$\cite{liu2022gate}, Te \cite{churchill2017toward}, ZrTe$_3$\cite{geremew2018current}}\label{be}
\end{figure}

We report in Figs.~\ref{be}a-c the distribution of the binding energies computed according to these three definitions.
In the first two cases in Figs.~\ref{be}a,b the energies follow a similar distribution with a sharp peak at low energies followed by a rather steep decline and long tail in the distribution. This behaviour is similar to what was previously observed for two-dimensional materials \cite{Mounet2018,campi2023}. It is also interesting to notice how the energy range for the second descriptor, Fig. \ref{be}b, is the same as the binding energies for area unit used for two-dimensional materials, which might hint towards an analogous applicability of this descriptor as in the case of two-dimensional materials, where we defined as easily exfoliable the materials with binding energies below 30 meV/\AA$^{2}$ and potentially exfoliable all the ones with binding energies included in the 30-120 meV/\AA$^{2}$ interval. 
In the figures, we mark the binding energies calculated according to our three definitions of nanowires with experimentally reported mechanical exfoliation, namely TiS$_3$\cite{island2015tise3,lipatov2018quasi}, MoI$_3$ \cite{teeter2024achieving}, Bi$_4$I$_4$\cite{liu2022gate}, Te \cite{churchill2017toward} and ZrTe$_3$\cite{geremew2018current}. TiS$_3$, Bi$_4$I$_4$, the tellurium chain and MoI$_3$ all lie below around 300 meV/\AA~ or 20 meV/\AA$^{2}$, suggesting indicators for the \emph{easily exfoliation} in 1D materials. On the other hand ZrTe$_3$, for which the exfoliation yielded ribbon-like structures with larger aspect ratios, exhibits higher binding energies (Figs. \ref{be}a,b), and could therefore fall within the category used for 2D materials\cite{mounet2018two} of \emph{potentially exfoliable} nanowires.

\begin{figure}[h!]
    \centering
    \begin{centering}
    \begin{subfigure}[c]{0.5\textwidth}
    \includegraphics[scale=0.82]{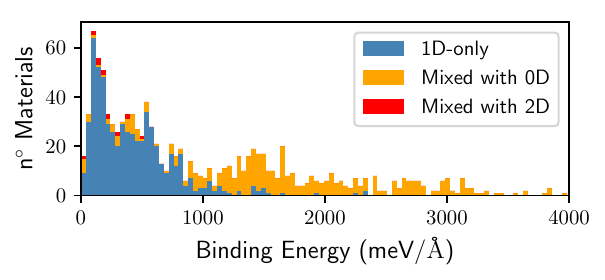}
    \caption{}\label{mixed2d} 
     \end{subfigure}%
    \begin{subfigure}[c]{0.5\textwidth}
    \includegraphics[scale=0.82]{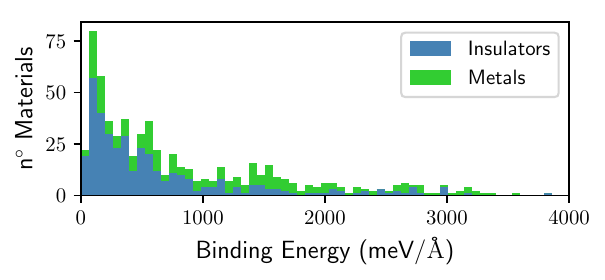}
    \caption{}\label{BE_bg}
    \end{subfigure}
    \end{centering}
    \caption{(a) Stacked histograms with the distribution of the binding energies per unit of length divided according to type of parent structure, formed by 1D-only wires, mixed with 0-D intercalants or mixed with 2D materials. (b) Distribution of the binding energies resolved for insulating and metallic compounds.}
\end{figure}

The third descriptor, accounting also for the wire stiffness shows instead a qualitatively different behaviour from the first two, with most materials distributed in a relatively narrow range and only few materials in the tail (See Fig. \ref{be}c). This means that a relatively large binding energy is generally compensated by a greater stiffness of the wire. This result is particularly interesting in light of the distribution shown in Fig.~\ref{mixed2d}, where we report the distribution of the binding energies divided by the characteristic geometry of the bulk parent, \emph{i.e.} composed of 1D wires only or 1D wires mixed with components of different dimensionality (2D layers or 0D intercalants).

Similarly to our previous results\cite{Mounet2018,campi2023} we can in fact note that the region of lower binding energies is largely dominated by materials whose parent 3D structure is composed exclusively of 1D substructures
held together by purely vdW forces, while moving towards higher binding energies, materials with zero-dimensional intercalated compounds become far more common, forming a second peak at higher energies and dominating the tail of the distribution. For these materials it is common to observe a charge transfer between intercalated 0D units and 1D wires and results in stronger, partially ionic bonds. It would be rather difficult to obtain these materials via a purely mechanical exfoliation but they are likely to be good candidates for a chemical exfoliation in solution with a solvent able to reduce the interaction between the 1D wires and the ionic intercalated components \cite{Bianco2013,Griffin2020, Coleman2011two,Coleman2011large}. This is especially true considering that the wires extracted from these materials are on average characterized by a higher stiffness (See Fig. \ref{be}c) that proved to be a key factor in two-dimensional materials to determine the quality of flakes obtained by liquid exfoliation\cite{Backes2019}. 

Finally, it is interesting to notice how the binding energies resolved according to the electronic properties of the one-dimensional material. Fig. \ref{BE_bg} shows a behaviour analogous to what observed in two-dimensional materials \cite{campi2023} with semiconducting or insulating materials being prevalent at low binding energy while metallic materials are more evenly distributed and in percentage dominating the higher end of the distribution.

\subsection{Electronic properties of the 1D nanowires}
\begin{figure}[h!]
    \centering
         \includegraphics[scale=1]{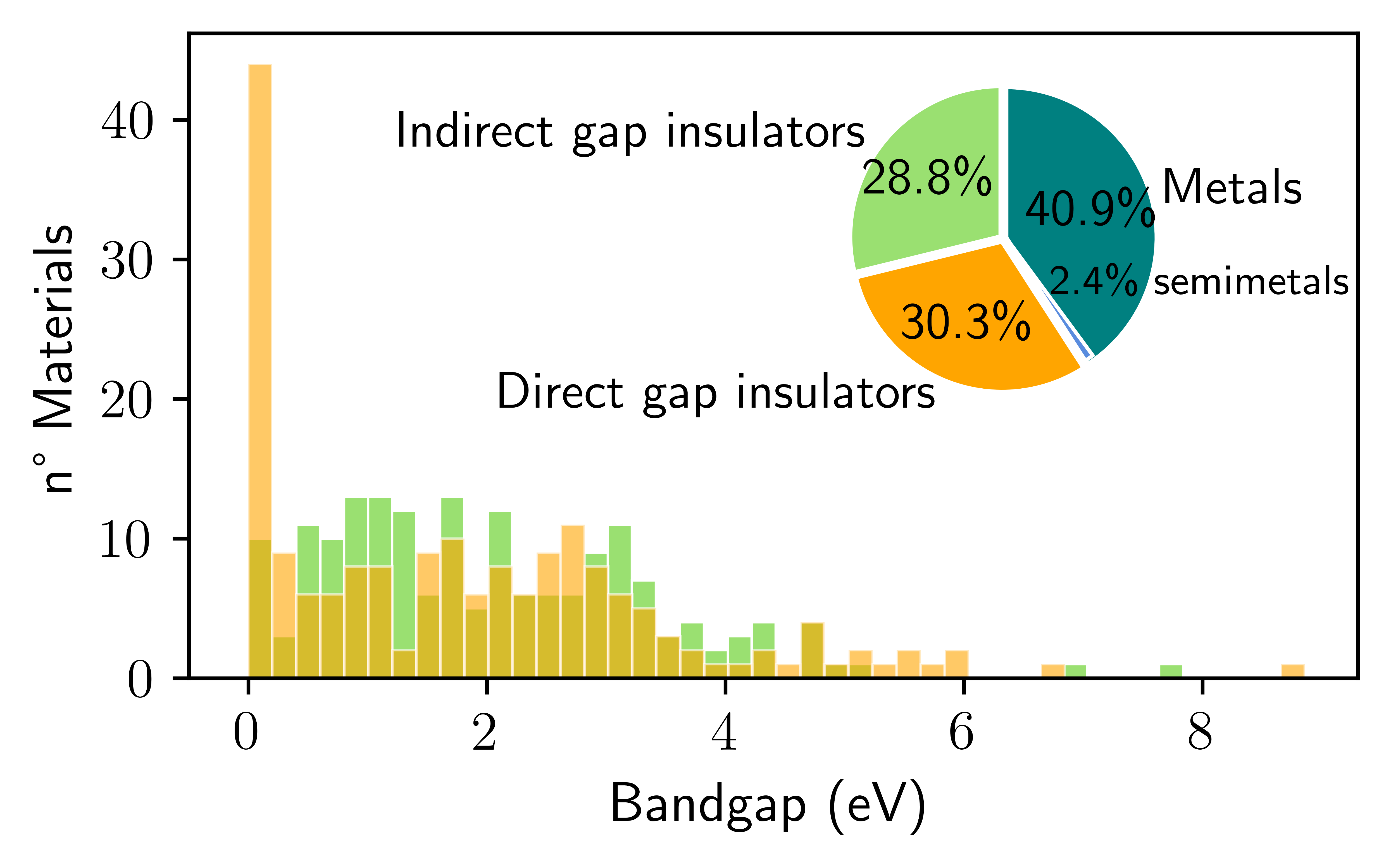}
    \caption{(a) Distribution of the band gaps for the insulating and semiconducting materials in
the database, divided by direct and indirect gap. Inset: percentage of metallic, direct and indirect semiconducting materials.}
    \label{BG_direct_indirect}
\end{figure}

The structural optimization proved to be successful for a total of 686, for all these materials we computed band structures at the PBE level along periodic paths.
We find that almost 41\% of the materials are metallic while the rest are insulating or semiconducting (see Fig. \ref{BG_direct_indirect}, inset). In the inset, we highlight as \emph{semimetals} 7 metallic materials where the bottom of the conduction band lies below the top of the valence band. 
In this work we have deliberately treated all structures as non-magnetic. Magnetic order in 1D is intrinsically fragile, since fluctuations always suppress long-range order so the fully ordered solutions provided by DFT are not representative of the true ground state. The largest deviations are expected for the wires containing 3d or 4f elements (about 10\% of the database), where magnetism could change the fraction of metallic systems and produce false positives among the charge density waves candidates (see next section). The effect on binding energies is instead expected to be negligible as observed for two-dimensional materials \cite{mounet2018two}. A detailed analysis of magnetism in 1D wires will be carried out in future studies.

From Fig. \ref{BG_direct_indirect}, the band gaps see a huge contribution for very low values, with almost sixty materials characterized by bandgaps smaller than 0.1 eV, about 9\% of the total materials. This result includes many materials with a single band crossing at the Fermi level with a Dirac-like dispersion, that are classified as extremely small bandgap materials either because of the physical presence of an avoided crossing or by the numerical error due to the finite size of the grid. With a more detailed analysis, we could identify 23 systems with a Dirac crossing at $\Gamma$ or $Z$, and four-fold degenerate eigenvalues at the crossing point. 11 of those are also dynamically stable  (see next section): P$_2$Pd$_2$S$_8$, P$_2$, N$_2$S$_2$, As$_2$Se$_2$, H$_4$Li$_2$O$_2$, As$_2$Te$_2$, As$_2$O$_2$, Sb$_2$Te$_2$, Na$_2$, Se$_4$Tl, O$_6$P$_2$. Other 7 show a linear-like dispersion around the Fermi energy, at k-point between $\Gamma$ and Z of the BZ, that could reveal a Dirac crossing with more dense k-grid calculations. 
These materials are particularly interesting both for topological phases that can emerge where the inclusion of spin-orbit coupling opens a non-trivial gap or for excitonic insulators where the excitons (likely highly energetic due to the lack of screening \cite{grillo2026quasiparticle}) can condensate in an exotic state\cite{Varsano2017}. 
We also identify 7 materials in which spin-orbit coupling (SOC) splits the Dirac-cone degeneracies, making them interesting for investigating possible Weyl semi-metallic behavior: Cu$_2$N$_2$, Cl$_6$Mo$_2$O$_2$, B$_2$Se$_6$, Cl$_8$Co$_2$, I$_2$O$_2$, B$_2$S$_6$ and Ag$_2$N$_2$. Of those, only the latter results dynamically unstable, while the others are either stable or $\Gamma$-unstable (see next section).
We provide the electronic band structure of all the interesting materials in the Supporting Information.
Lastly, around 0.2\% of the total number of wires are semiconductors with an extremely small bandgap ($\leq $ 0.1 eV).

For larger gaps, the distribution is quite uniform up to 3 eV energy range, after which it decays quite rapidly. All these materials could have a great potential impact for the realization of ultrascaled transistors with all-around dielectric control or in electro- or photo-catalysis where one-dimensional materials can offer the ultimate surface-to-volume ratio.

The proportion between direct and indirect gap of semiconducting and insulating wires in Fig. \ref{BG_direct_indirect} is well balanced, with a weak prevalence of indirect bad gaps between 0.2 eV and 2 eV. Materials with band gap smaller than 0.2 eV have instead a predominance of direct gaps, among which, however, potential Dirac-cone materials are included.

The one-dimensional database contains 44 systems with a Mexican-hat valence band structure, of which 50\% dynamically stable. The majority of them contain heavy elements like Sn, Sb, Br and Bi. Mexican hat dispersions induce a degeneracy at the Fermi surface, leading to an enhancement of the density-of-state. This effect is usually exploited in thermoelectric applications, where it could boost the thermoelectric figure of merit ZT and the Seebeck coefficient\cite{wickramaratne2015electronic,chen2018large}.
In the Supporting Information the electronic band structure of these materials are reported.

\begin{figure}[h!]
    \centering
        \includegraphics[scale=1]{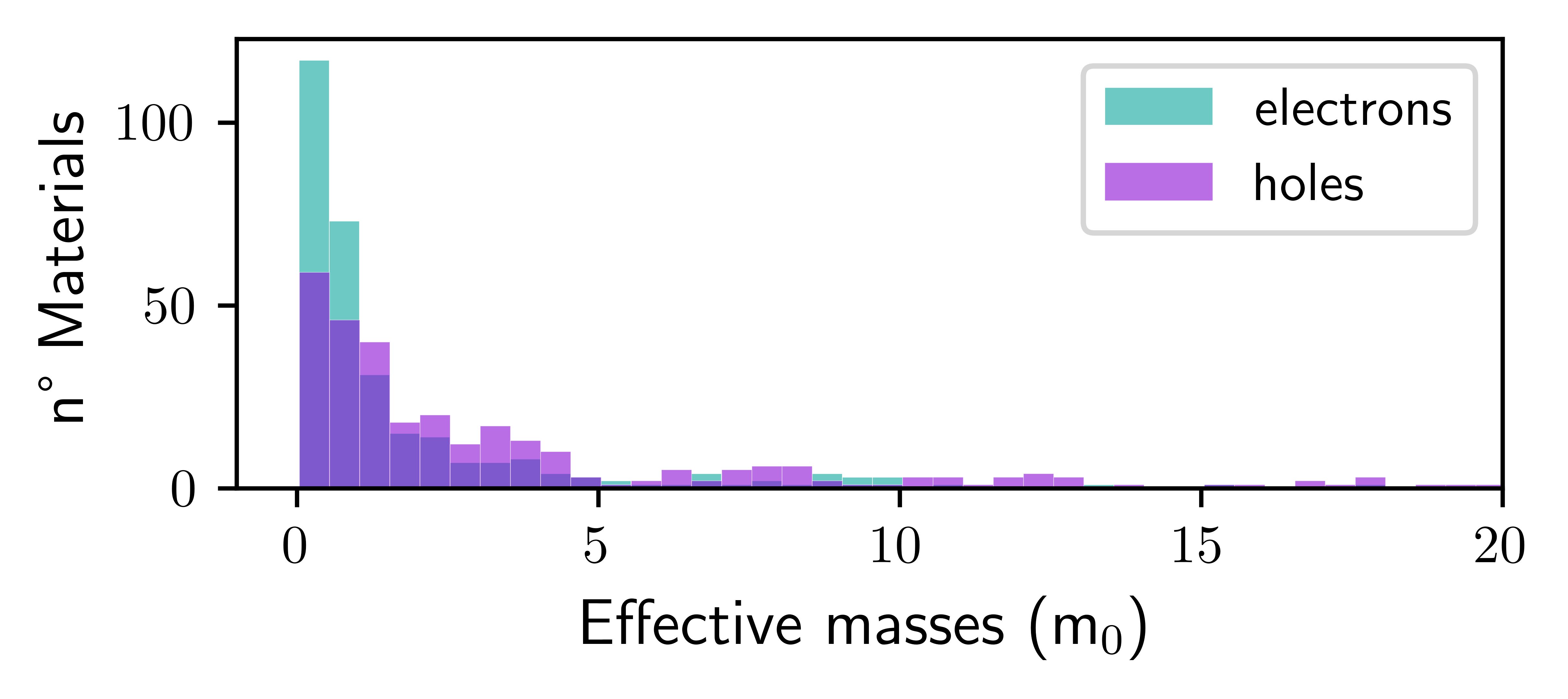}
        \caption{Distribution of the effective masses for insulating and semiconducting materials in the database (E$_g\geq$0.1 eV). We show $m^*<$ 20 $m_0$ only, that represent $\sim$ 95 \% of the 1D materials.} \label{parabolic_effective_masses}
\end{figure} 

From the band structures, we calculate the effective masses of the 1D wires, which quantifies the curvature of the electronic bands and therefore how strongly the bands disperse along the periodic direction.
We compute the hole and electron effective masses by a parabolic fit of the band at the valence band maximum and conduction band minimum, respectively. 
The distribution of the effective masses in Fig. \ref{parabolic_effective_masses} shows that most of the systems have 
effective masses lower than 5 $m_0$ for both holes and electrons, corresponding to about 74\% and 88\% of the semiconducting wires, respectively. Within this range, a clear peak is observed at $m^*/m_0$ lower than 1. 
These materials can be particularly interesting for electronic application in downscaled devices, as the electron mobility is inversely proportional to the effective mass of the material.
In contrast, around 5\% of all systems present a large effective mass greater than 20 $m_0$ (not shown in the plot). 

\begin{figure}[h!]
    \centering
    \begin{centering}
    \begin{subfigure}[c]{0.5\textwidth}
    \includegraphics[scale=0.80]{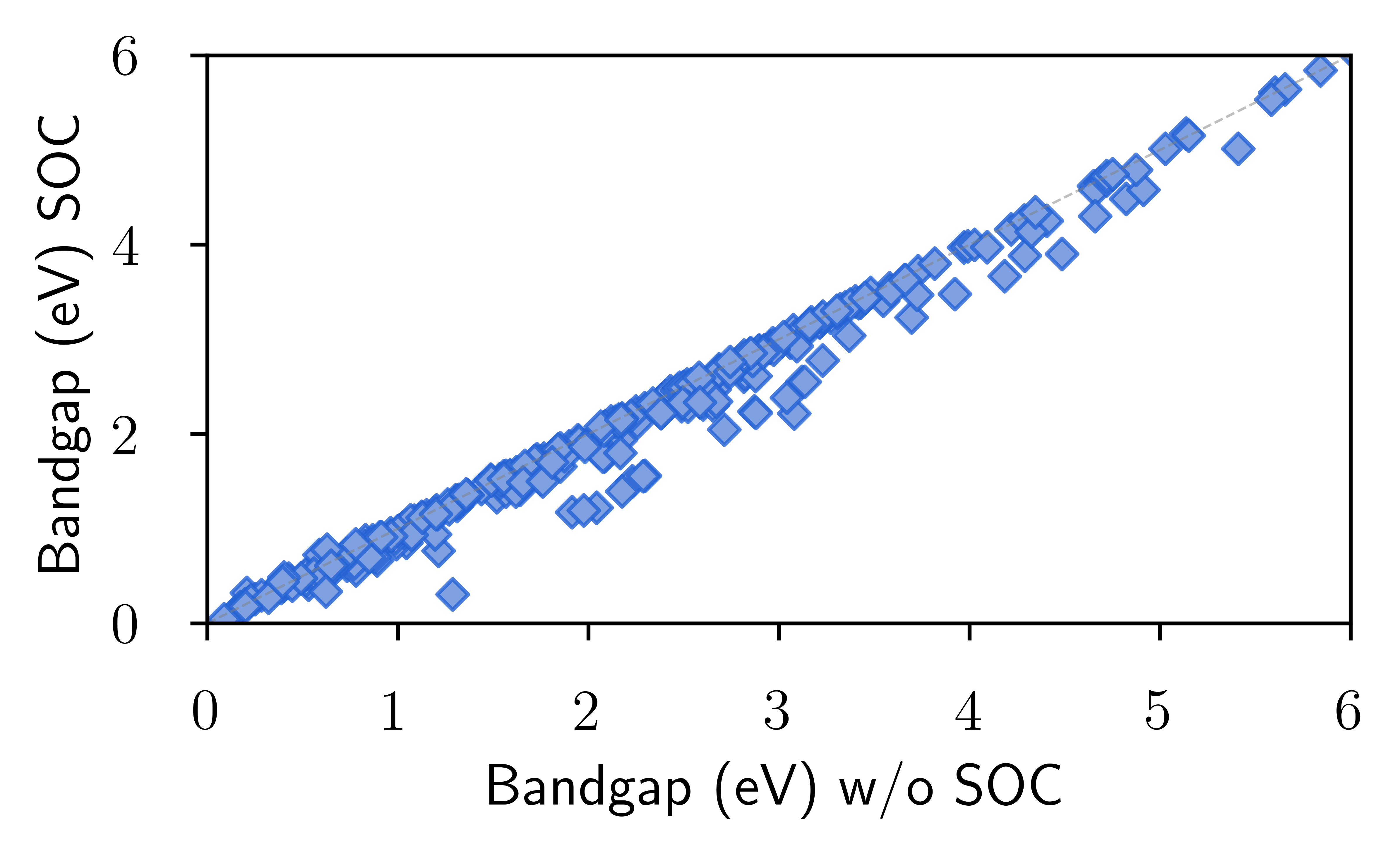}
   \caption{}
    \label{BG_soc_nosoc} 
    \end{subfigure} %
    \begin{subfigure}[c]{0.5\textwidth}
    \includegraphics[scale=0.80]{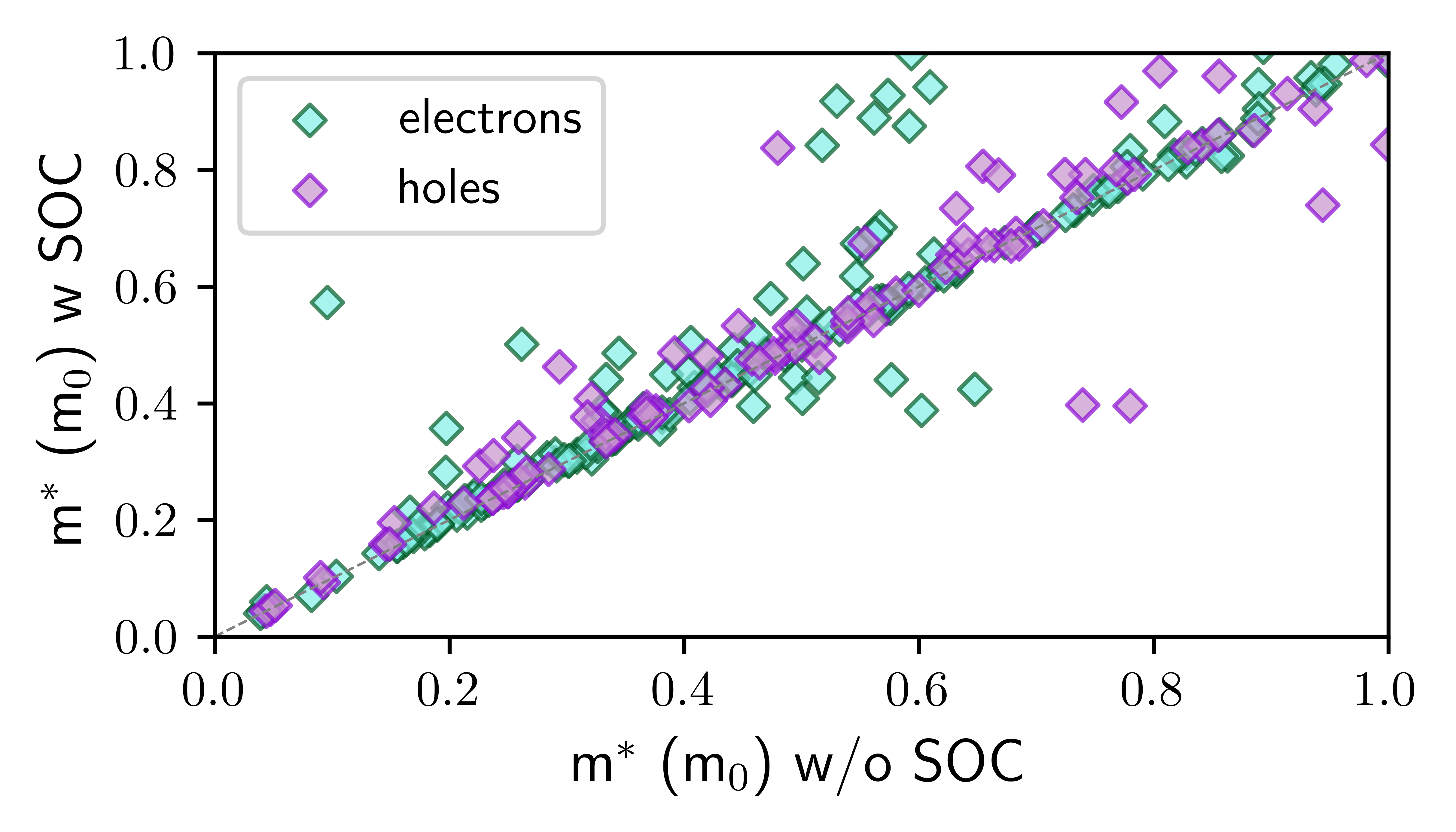}
   \caption{}
    \label{mass_soc_nosoc} 
\end{subfigure}
\end{centering}
\caption{(a) Band gaps and (b) effective masses comparison with and without the inclusion of spin-orbit coupling (zoom on $m^*<$1 $m_0$).}
\end{figure}

Finally, we compute band structures at the PBE level including the spin-orbit coupling (SOC) of the 686 materials in the one-dimensional database. In Fig. \ref{BG_soc_nosoc} the bang gaps with and without SOC are compared for the insulating and semiconducting materials (E$_g$ $\geq$ 0.1 eV). We notice that the band gaps remain unchanged by the inclusion of SOC for most of the systems. In cases where it changes, the SOC-induced band splitting mostly leads to a reduction of the gap rather than an increase. The largest changes occur for wires containing heavier elements. 
In Figure \ref{mass_soc_nosoc} the hole and electron effective masses are computed with and without the inclusion of spin-orbit coupling. As before, a strong SOC effect is mostly restricted to materials with heavier elements only, such as Sb, Te, I, Tl, Pb or Bi.

We note that the weak electronic screening in one-dimensional materials may lead to strong renormalization of their band gaps, which are expected to increase when these effects are properly accounted for using hybrid functionals or the GW approximation \cite{grillo2026quasiparticle}.

\subsection{Dynamical stability of the 1D nanowires}
Free-standing one-dimensional materials can be quite fragile. Ideal metallic 1D materials, in particular, are known to be subject to dynamical instabilities arising from the divergence of the electronic response at the \emph{nesting} wavevectors, which connect two points at the Fermi level. This divergence is reflected in a sharp anomaly in the phonon dispersion \cite{kohn1959image,peierls1955quantum,pouget2016peierls}. For these reasons, to establish whether or not a material can exist as an isolated freestanding wire it is essential to evaluate its dynamical stability by computing the full phonon dispersion along the periodic direction. Negative modes in the phonon dispersion signal the insurgence of a dynamical instability, and their type and location is useful to understand the origin as well as the possible outcome of such instabilities.  
\begin{figure}[h!]
    \centering
    \includegraphics[scale=1.0]{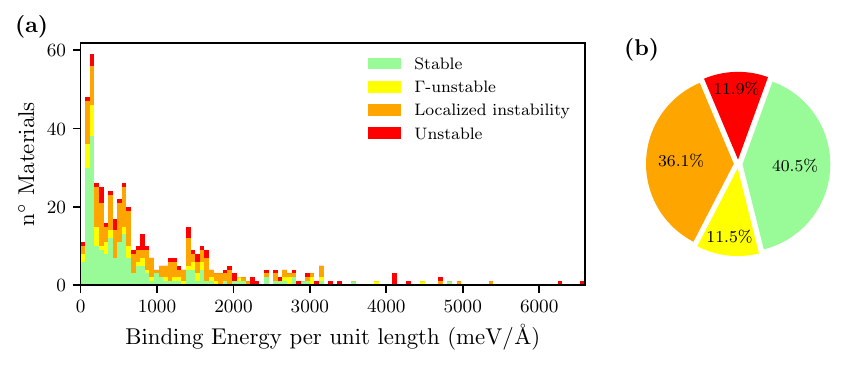}
    \caption{(a) Distribution of the binding energies for dynamically stable compounds, compounds showing instability at the $\Gamma$ point, compounds showing localized instability and unstable compounds. (b)  Percentage of stable, $\Gamma$-unstable, locally unstable and generally unstable materials.} \label{BEphdist}
\end{figure}
In Fig. \ref{BEphdist} we classify our one-dimensional materials into four categories: i) stable materials, \emph{i.e.}, materials that do not show any significant form of instability in the phonon dispersion, with a sensibility of $\pm$ 10 cm$^{-1}$. ii) $\Gamma$-unstable materials, that show instabilities at the $\Gamma$-point of the Brillouin zone. iii) Materials that show an instability involving one or a few modes localized in a relatively narrow region of the 1D Brillouin zone, resembling a charge-density wave-like behavior. 
Finally iv) unstable compounds characterized by diffuse instabilities involving multiple modes over extended regions of the Brillouin zone. 

In our analysis we found that 40.5\% of the materials for which we successfully obtained a converged phonon dispersion are stable in the form as-extracted from the three-dimensional parent. Unsurprisingly most of these materials are associated with low binding energies (See Fig. \ref{BEphdist}) and usually a semiconducting ground state. 11.5\% of materials show an instability at the $\Gamma$-point of the Brillouin Zone instead. These materials are likely to have a stable ground state with the same unit-cell periodicity, which can be reached by breaking the enforced symmetry through distortion along the unstable modes \cite{cignarella2024searching}. In fact, these instabilities are likely due to a symmetry inherited from the three dimensional bulk material that the structural optimization algorithm preserves during the relaxation.
As shown in Fig. \ref{BEphdist}, also these wires are mostly associated with low-binding energy materials. 
On the other hand,  materials with localized instabilities are distributed more uniformly, 
following more closely the distribution of metallic materials (see Fig. \ref{BE_bg}), for which this behaviour is by far the most common (the 37\% of the metallic systems against 22\% of insulating ones). In these cases, a stable free-standing one-dimensional phase of the material may exist, but finding it requires considering a supercell commensurate to the point in the Brillouin Zone where the instability appears \cite{cignarella2024searching}. The stable structure would be similar to the one presented in this study, as can be derived from the high-symmetry structure by imposing a small periodic distortion. 
In both cases of $\Gamma$-unstable and materials with localized instabilities, 
the symmetric phase present in our database may exist as a metastable phase or high-temperature phase, when anharmonic effects are taken into account \cite{cignarella2025extreme}.
Moreover, metallic CDW-like unstable wires could represent an ideal playground for applications involving charge density wave physics \cite{BALANDIN202274,Geremew2018,Park2020}.
Finally, we find that 11.9\% of the materials are generally unstable, with multiple and diffuse instabilities in their phonon dispersion making them unlikely to exist as one-dimensional materials once isolated from their environment.     
This study highlights how important is to carefully analyze the entire phonon spectrum when assessing the dynamical stability of one-dimensional wires, while also uncovering interesting forms of instabilities, like those suggesting charge-density-waves formation.

\section{Conclusions}
In conclusion, we screened three large experimental databases of crystalline structures in search of inorganic one-dimensional materials that could be isolated from weakly bonded three-dimensional parents. We identified 686 wires and proposed three different metrics to describe the likelihood of their experimental isolation. We studied their electronic properties revealing that isolated 1D systems possess a broad range of electronic features of interest from both theoretical and application perspectives, including Dirac cones, Mexican-hat bands, possible Weyl-semimetal phases, and low effective masses, which could be exploited in dangling-bond-free channels in next-generation downscaled devices.  
Crucially we studied their dynamical stability based on a full phonon dispersion, which allowed us to identify stable materials as well as interesting forms of instability like charge density wave behaviours. Notably, 52\%  of the wires in our database are likely to be stable in their isolated, freestanding phase.

One-dimensional materials could offer new perspectives for both fundamental research and device applications.
These results can provide useful guidance for experimentalists and theorists alike, offering a way to identify the materials more likely to be successfully isolated as well as pinpointing the interesting properties and behaviours of exfoliable one-dimensional materials.
With this work, we provide a large, openly accessible database of materials and their properties to support further exploration of the largely unexplored realm of 1D wires.

\section{Data availability}
All the 686 materials and their properties will be available in the MC1D Discover section of the Materials Cloud together with the full AiiDA database in the Archive section and allowing also for a full exploration of the provenance graph in the Explore section. In the Discover section, for all material is given a comprehensive summary of the properties including the optimized structure, the electronic bands (with and without SOC), the effective masses, the binding energies and information on the parent 3D material. 
\section{Methods}\label{methods}
We use the Quantum ESPRESSO \cite{Giannozzi2009,Giannozzi2017} package with the SSSP PBE efficiency pseudopotentials library (version 1.0) \cite{Prandini2018}, a library of carefully tested pseudopotentials from different sources \cite{VANSETTEN2018,TOPSAKAL2014,DALCORSO2014,Willand2013,SCHLIPF2015,GARRITY2014}. For each material, the wavefunction and charge-density cutoffs are chosen as the highest suggested by the SSSP library for all the elements in the compound. For bulk materials, we use two properly vdW-corrected functionals, namely the vdW-DF2 functional \cite{lee2010} with C09 exchange \cite{cooper2010,hamada2010} (DF2-C09). Sampling of the Brillouin zone is performed using a $\Gamma$-centred Monkhorst-Pack grid \cite{MPgrid}, with the smallest number of k-points in each direction of the reciprocal lattice guaranteeing a spacing of at least to 0.2 \AA$^{-1}$. All relaxations and binding energies of 3D structures are computed using a Marzari-Vanderbilt-DeVita-Payne cold smearing \cite{MVsmearing} of 0.02 Ry. The volume is optimized letting all the cell vectors and angles free to move until the pressure is below 0.5 kbar and the internal forces are lower than 10$^{-3}$ a.u.
When optimizing and computing the electronic properties of isolated 1D units, a vacuum space of at least 20 \AA\ is used along the non-periodic directions and a variable cell relaxation is performed only along the periodic direction using the PBE\cite{PBE} functional with the same pseudopotentials and equivalent k-points sampling.  
The quantum volume V$_0$ is computed as implemented in the Quantum ENVIRON package\citep{andreussi2012revised}, taking the linear coefficient of the PV energy term, placing the system in a vacuum environment (with static permittivity $\varepsilon$=1.0000001).
We obtain the Young's modulus by fitting the energy-strain curve along the periodic direction with a quadratic function \cite{cignarella2024searching}. 
Band structures are computed along high-symmetry paths with a k-points density of 0.1 \AA$^{-1}$. All the structures, even the ones with atoms that might allow a magnetically ordered ground state, are treated as non magnetic in a spin-unpolarized approximation.
SOC calculations are performed using the PBE functional and fully-relativistic pseudopotentials from PseudoDojo library v0.4 \cite{van2018pseudodojo}, k-points distance in the periodic direction of 0.1 \AA$^{-1}$and k-points grid for the band structure calculations of 1$\times$1$\times$100.
Phonon band dispersion are computed using density-functional-perturbation theory (DFPT)\cite{baroni2001phonons}, with a q-point distance of 0.3 \AA$^{-1}$. For phonon dispersions displayed in the Materials Cloud Explore section, we employ the acoustic sum rule implemented in Ref. \citenum{lin2022general} that imposes the full Born-Huang invariance conditions to correct long-wavelength behaviour of the acoustic phonons.

\begin{acknowledgement}
This research was
supported by the NCCR MARVEL, a National Centre of Competence in 
Research, funded by the Swiss National Science Foundation (grant number 
205602).
C.C. and N.M acknowledge also support from the Deutsche Forschungsgemeinschaft (DFG) under Germany’s Excellence Strategy (EXC 2077, No. 390741603, University Allowance, University of Bremen) and MaX “Materials Design at the Exascale”, a Center of Excellence co-funded by the European High Performance Computing Joint Undertaking (JU) and participating countries under grant agreement No. 101093374, and Lucio Colombi Ciacchi, the host of the “U Bremen Excellence Chair Program”.
C.C. acknowledges the Swiss National Science Foundation for funding under the Postdoc.Mobility program (grant number P500-2\_235339).
D.C. acknowledges the CINECA allocations under the ISCRA initiative and the UNIMIB-CINECA agreement, for the availability of high-performance computing resources and support.

Conflict of Interest: The authors declare no competing financial interest.
\end{acknowledgement}

\bibliography{library}

\end{document}